\documentclass[pra,twocolumn,showpacs]{revtex4}
\usepackage{amsmath}
\usepackage{amssymb}
\input{epsf}
\usepackage{epsfig}

\newcommand{\rem}[1]{}

\begin{document}
       
\title{Classical versus quantum errors in quantum computation of
dynamical systems}
\author{Davide Rossini}
\email{d.rossini@sns.it}
\affiliation{Center for Nonlinear and Complex Systems, Universit\`a degli 
Studi dell'Insubria}
\affiliation{NEST- INFM \& Scuola Normale Superiore, Piazza dei Cavalieri 7,
56126 Pisa, Italy}
\author{Giuliano Benenti}
\email{giuliano.benenti@uninsubria.it}
\author{Giulio Casati}
\email{giulio.casati@uninsubria.it}
\homepage{http://www.unico.it/~dysco}
\affiliation{Center for Nonlinear and Complex Systems, Universit\`a degli 
Studi dell'Insubria and Istituto Nazionale per la Fisica della Materia, 
Unit\`a di Como, Via Valleggio 11, 22100 Como, Italy}
\date{\today}

\begin{abstract} 
We analyze the stability of a quantum algorithm simulating the 
quantum dynamics of a system with different regimes, ranging 
from global chaos to integrability.
We compare, in these different regimes, the behavior of the 
fidelity of quantum motion when the system's parameters are 
perturbed or when there are unitary errors in the quantum
gates implementing the quantum algorithm. While the first kind
of errors has a classical limit, the second one has no classical
analogue. It is shown that, whereas in the first case
(``classical errors'') the decay of fidelity is very 
sensitive to the dynamical regime, in the second case
(``quantum errors'') it is almost independent 
of the dynamical behavior of the simulated system. 
Therefore, the rich variety of behaviors found in the study
of the stability of quantum motion under ``classical'' perturbations 
has no correspondence in the fidelity of quantum computation under 
its natural perturbations.
In particular,
in this latter case it is not possible to recover the semiclassical
regime in which the fidelity decays with a rate given by the 
classical Lyapunov exponent.
\end{abstract}
\maketitle

\section{Introduction} \label{sec.1}

Fidelity is a very convenient tool to characterize the stability
of quantum computation. 
It is defined as  
$f(t)=\vert \langle \psi(t)  \vert \psi_\epsilon(t)\rangle \vert^2$,
where the two state vectors $\vert \psi(t)\rangle$ and 
$\vert \psi_\epsilon(t)\rangle$ are obtained by evolving the 
same initial state $\vert \psi_0\rangle$, under ideal
or imperfect quantum gates, respectively. Here $\epsilon$
measures the imperfection strength and we assume that the 
perturbed gates are still unitary.
If the fidelity is close to one, the results of  
the quantum computation are close to the ideal ones, while,
if $f$ is significantly smaller than one, then quantum
computation does not provide reliable results. 

More generally the fidelity (also called Loschmidt echo) is a quantity
of central interest in the study of the stability of 
dynamical systems under perturbations
\cite{peres,jalabert,pastawski,jacquod,tomsovic,felix,PRE,cohen,prosen,lloyd,baowenli,mirlin,marcos,heller,zurek,geisel,veble,wenge}.
The decay of fidelity
in time exhibits a rich variety of different behaviors, 
from Gaussian to exponential or power-law decay, depending,
e.g., on the chaotic or integrable nature of the system under 
investigation, on the initial state (coherent state, mixture, etc.),
and, for integrable systems, on the shape of the perturbation
and on initial conditions. In particular, in the chaotic,
semiclassical regime, and for strong enough perturbations, it has 
been shown that the decay rate is perturbation independent and 
determined by the Kolmogorov-Sinai entropy, related to the 
Lyapunov exponent of classical chaotic dynamics
\cite{jalabert}. 

On the other hand, the simulation 
of the quantum dynamics of models describing 
the evolution of complex systems promises to become the first 
application in which a quantum computer with only a few tens
of qubits may outperform a classical computer. Indeed, efficient 
quantum algorithms simulating the quantum evolution of 
dynamical systems like the baker's map \cite{schack}, 
the kicked rotator \cite{bertrand}, and the sawtooth map 
\cite{simone} have been found, and important physical 
quantities could be extracted from these models already with 
less than 10 qubits \cite{simone2,pomeransky,libro}.
Therefore, these quantum algorithms may constitute the ideal 
software for short- and medium-term quantum computers operating
with a small number of qubits and the most suitable testing 
ground for investigating the limits to quantum computation
due to imperfections and decoherence effects.
In this context, we point out that the fidelity of 
quantum computation has been evaluated for 
the quantum baker's map using a three-qubit NMR-based
quantum processor \cite{cory}. 
We also note that efficient quantum algorithms to compute 
the fidelity have been proposed in Refs.~\cite{emerson,laflamme}.

From the viewpoint of computational complexity, the following question
naturally arises: given a generic dynamical system, is it possible to
find its solution at time $t$ efficiently, including into consideration
unavoidable computational errors?
We recall that the classical dynamics of chaotic systems is characterized 
by exponential sensitivity: any amount of error 
in determining the initial conditions diverges exponentially, 
with rate given by the largest Lyapunov exponent $\lambda$. 
This means that, when following a given orbit, one digit of 
accuracy is lost per suitably chosen unit of time.
Therefore, to be able to follow one orbit up to time $t$
accurately, we must input $O(t)$ bits of information
to determine initial conditions.
On the other hand, the orbit of a non-chaotic system is
much easier to simulate, since errors only grow 
linearly with time.
Owing to the exponential instability, classical chaotic dynamics is 
in practice irreversible, as shown by Loschmidt echo numerical
simulations of Ref.~\cite{casati}:
if, starting from a given classical distribution in phase space, 
we simulate the dynamical evolution up to time $t$ and then, 
by inverting at time $t$ all the momenta, we follow the backward 
evolution, we do not recover the initial distribution at time $2t$. 
This is because any amount of numerical error in computer
simulations rapidly effaces the memory of the initial conditions.
On the contrary, the same numerical simulations in the quantum
case show that time reversibility is preserved in the
presence of small errors. 

In view of the above considerations, it is natural to inquire
whether the degree of stability of a quantum algorithm
depends on the nature (chaotic or non-chaotic) of the simulated 
dynamics. 
We will show that the decay of the fidelity of a quantum algorithm
in the presence of perturbations in the quantum gates 
is almost independent of the dynamical behavior of the simulated system.
 
In this paper, we will consider
a quantum system, the so-called sawtooth map, which can be 
simulated efficiently on a quantum computer and whose underlying 
classical dynamics, depending on system's parameters, can be chaotic 
or non-chaotic. 
We will outline the main differences that 
occur in calculating the fidelity decay with ``classical'' and 
``quantum'' perturbations on the dynamical system.
\begin{itemize}
\item
By \emph{classical perturbations}, we mean perturbations of the
system's parameters that have a classical limit.
For instance, in this paper we perturb, at each map step 
of the sawtooth model, the kicking strength $k$ 
by a small amount $\delta k (t)\ll k$, where $t$ measures
the number of map iterations. 
Note that this kind of perturbation, when applied to the 
classical motion, disturbs a given orbit by a small amount 
at each map step and therefore, to some extent, mimics
the presence of round-off errors in a classical computer.
\item
By \emph{quantum perturbations}, we mean errors introduced at each quantum 
gate (in this paper, we consider unitary, memoryless errors). 
These quantum errors are unavoidable during a quantum computation, 
due to the imperfect control of the quantum computer hardware and 
they do not have classical analogue. We will show 
that the fidelity 
decay evaluated with quantum errors is not capable of
distinguishing between the classically integrable or chaotic 
nature of the simulated dynamics, being essentially independent 
of it.
\end{itemize}

This paper is organized as follows. In Sec.~\ref{sec.2}, we briefly 
describe the sawtooth map model and a quantum algorithm
which efficiently simulates it. 
We also introduce our quantum and classical error models and 
discuss how to efficiently evaluate the fidelity on 
a quantum computer.
In Sec.~\ref{sec.3}, based on extensive numerical 
simulations, we analyze the differences between the 
fidelity decay in the presence of classical and
quantum error.
Finally, in Sec.~\ref{sec.4} we present our conclusions.

\section{The perturbed quantum sawtooth map model} \label{sec.2}

In order to illustrate the striking differences between
the fidelity decays induced by classical and quantum errors, 
here we consider the quantum sawtooth map model. 
This map is one of the most extensively studied dynamical systems,
since it exhibits a rich variety of different dynamical regimes, 
ranging from integrability to chaos, and interesting physical
phenomena like normal and anomalous diffusion, dynamical 
localization, and cantori localization
\cite{percival,fausto,fausto2,stadium}. 

\begin{figure}
\centerline{\epsfxsize=8.2cm \epsffile{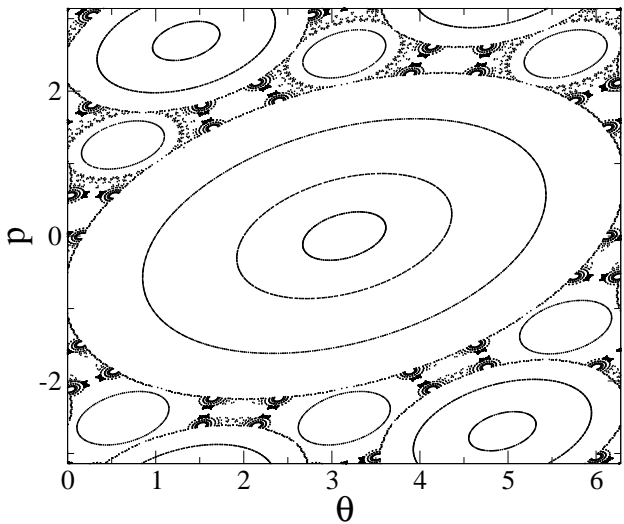}}
\caption{Poincar\'e cross sections for the classical sawtooth map
in the quasi-integrable regime at $K=-0.5$.
We show $7$ trajectories inside the integrable islands
and a single trajectory filling the anomalously diffusive 
region.}
\label{fig1}
\end{figure}

The sawtooth map is a periodically driven dynamical system, described by the
Hamiltonian
\begin{equation}
H(\theta,n,t) = \frac{n^{2}}{2} - \frac{k (\theta - \pi)^{2}}{2}
\sum_{j=-\infty}^{+\infty} \delta(t-jT),
\label{hammap} \end{equation}
where $(n,\theta)$ are the conjugated action-angle variables
($0 \leq \theta < 2\pi$).
The time evolution $t \to t+T$ of this system is classically described by the
map 
\begin{equation}
\bar{n} = n + k (\theta - \pi), \quad \bar{\theta} = \theta + T \bar{n},
\label{hammap2}
\end{equation}
where the bars denote the variables after one map iteration.
By rescaling $n \to p=Tn$, the classical dynamics is seen to depend
only on the parameter $K=kT$. 
The classical motion is stable for $-4 \leq K \leq 0$ and 
completely chaotic for $K<-4$ and $K>0$: the maximum Lyapunov exponent 
is $\lambda=\ln[(2+K+\sqrt{K^2+4K})/2]$ for $K>0$, 
$\lambda=\ln|(2+K-\sqrt{K^2+4K})/2|$ for $K<-4$, and 
$\lambda=0$ for $-4\le K \le 0$. 
As shown in Fig.~\ref{fig1}, in the stable, quasi-integrable 
regime, the phase space 
has a complex structure of elliptic islands down to smaller and 
smaller scales. Note that the integrable islands are surrounded by
a non-integrable region, and that each trajectory diffuses
(anomalously) in this region.
The cases $K=0,-1,-2,-3,-4$ are integrable.

The quantum evolution in one map iteration is described by 
the unitary operator $\hat{U}$:
\begin{equation}
\vert \bar{\psi} \rangle = \hat{U} \vert \psi \rangle =
e^{-iT\hat{n}^{2}/2} \, e^{ik(\hat{\theta} -\pi)^{2}/2} 
\vert \psi \rangle,
\label{quantmap} 
\end{equation}
where $[\hat{\theta},\hat{n}]=i$,
$\hat{n} = -i \partial/\partial \theta$, 
and $\vert \psi(\theta + 2\pi) \rangle = 
\vert \psi(\theta) \rangle$. Note that we have set $\hbar=1$.
We study this map on the torus 
$0 \leq \theta <2\pi$, $-\pi \leq p < \pi$.
The effective Planck constant is given by $\hbar_\text{eff}=T$.
Indeed, if we consider the operator $\hat{p}=T\hat{n}$
($\hat{p}$ is the quantization of the classical rescaled 
action $p$), we have
\begin{equation}
[\hat{\theta},\hat{p}]=T[\hat{\theta},\hat{n}]=i T =i 
\hbar_\text{eff}.
\end{equation}
The classical limit $\hbar_\text{eff}\to 0$ is obtained by taking
$k\to\infty$ and $T\to0$, while keeping $K=kT$ constant.
We consider Hilbert spaces of dimension $N=2^{n_q}$, where
$n_q$ is the number of qubits, and set $T=2\pi/N$.
Therefore, $\hbar_{\text{eff}}\propto 1/N=1/2^{n_q}$ 
drops to zero exponentially with the number of qubits.

The operator $\hat{U}$ can be written as the product 
of two operators, 
$\hat{U}_{k}= e^{ik(\hat{\theta}-\pi)^{2}/2}$
and $\hat{U}_{T}=e^{-iT\hat{n}^{2}/2}$.
Since $\hat{U}_k$ is diagonal in the $\theta$ representation,
while $\hat{U}_T$ is diagonal in the $n$ representation,
the most convenient way to simulate map (\ref{quantmap}) on a 
classical computer is based on the forward-backward fast Fourier 
transform between $\theta$ and $n$ representations, and requires
$O(N\log N)$ operations per map iteration.
The quantum computation takes advantage of the quantum 
Fourier transform and needs 
$O((\log N)^2)$ one- and two-qubit gates
to accomplish the same task \cite{simone,simone2}. 
More precisely, it needs $2n_q$ Hadamard gates and 
$3n_q^2-n_q$ controlled-phase shift gates.
Therefore, the resources required to the quantum 
computer to simulate the evolution of the sawtooth map are 
only logarithmic in the system size $N$, and there is an
exponential speed up, as compared to the best known 
classical computation.

Any experimental realization of a quantum computer has to face
the problem of errors, which inevitably set limitations to
the accuracy of the implemented algorithms. 
These errors can be due to unwanted couplings with the 
environment or to imperfections in the quantum hardware.
In this paper, we limit ourselves to consider 
\emph{unitary errors}, modeled by \emph{noisy gates}.
Such noise results from the imperfect control of the 
quantum computer.
For instance, in a NMR quantum computer the logic gates on 
qubits are simulated by applying magnetic fields to the  
system. 
If the direction or the intensity of the fields are 
not correct, a slightly different gate is applied, though it 
remains unitary. In ion-trap quantum processors, laser pulses are 
used to implement sequences of quantum gates \cite{blatt}.
Fluctuations in the duration of each pulse induce unitary errors, 
which accumulate during a quantum computation. 

As we have stated above, the implementation of the quantum algorithm
for the sawtooth map requires controlled-phase shift and 
Hadamard gates \cite{simone,simone2}. 
We choose to perturb them as follows. 
Controlled-phase shift gates are diagonal in the computational
basis and act non-trivially only on the four-dimensional Hilbert
subspace spanned by two qubits. In this subspace, we write each 
controlled-phase shift gate as 
$\tilde{C}={\cal E} C$, where $C$ is the ideal gate and 
the diagonal perturbation ${\cal E}$ is given by
${\cal E}={\rm diag}(e^{i\epsilon_0},e^{i\epsilon_1},
e^{i\epsilon_2},e^{i\epsilon_3})$. Therefore, the unitary error
operator ${\cal E}$ introduces unwanted phases.
The Hadamard gate can be seen as a rotation of the Bloch sphere
through an angle $\delta=\pi$ about the axis 
$\hat{u}_0=(\sin\theta_0\cos\phi_0,\sin\theta_0\sin\phi_0,
\cos\theta_0)$, where $\theta_0=\pi/4$ and $\phi_0=0$, so that 
$\hat{u}_0=(1/\sqrt{2},0,1/\sqrt{2})$. Since each one-qubit gate
can be seen as a rotation about some axis $\hat{u}$, 
unitary errors tilt the rotation angle:
$\hat{u_0}\to \hat{u}=
(\sin\theta\cos\phi,\sin\theta\sin\phi,\cos\theta)$, where 
$\theta=\theta_0+\nu_1$ and $\phi=\phi_0+\nu_2$.
We assume that the dephasing parameters $\epsilon_i,\nu_j$ 
($i=1,...,4$, $j=1,2$) are randomly and uniformly distributed
in the interval $[-\epsilon,\epsilon]$.
We also assume that the errors affecting different quantum 
gates are completely {\it uncorrelated}: every time we apply
a noisy gate, the dephasing parameters randomly fluctuate in
the (fixed) interval $[-\epsilon,+\epsilon]$. 
We note that the memoryless unitary error model has been 
widely investigated in the literature, see, e.g., 
Refs.~\cite{ciraczoller,pazzurek,song,biham,bettelli}.

We will compare the effect of noisy gates (''quantum errors'') 
with that of \emph{randomly fluctuating perturbations}
in the system's parameters (``classical errors'').
Here we choose to perturb the kicking strength $k$ in Eq.~(\ref{hammap2}) 
as follows: at each map step, $k$
is slightly changed by a small amount $\delta k(t)$, 
which is randomly chosen in the interval
$[- \delta k, \delta k]$.
Consequently, $\delta K(t)\equiv T \delta k(t) \in 
[-\delta K,+\delta K]$, where $\delta K \equiv T \delta k$.
As we have discussed in the introduction, this perturbation
models, to some extent, the effect of round-off errors 
in classical computation.

We will consider the following initial conditions:
\begin{itemize}
\item
A \emph{coherent Gaussian wave packet}:
\begin{equation} \vert \psi_0 \rangle_{\scriptscriptstyle G} = A
\sum_{n=0}^{N-1} e^{\frac{-(n - n_{0} )^{2}}{2 \sigma^{2}} + 
i \left(n - \frac{n_0}{2}\right)
\theta_0} \vert n \rangle, 
\label{gauss}
\end{equation}
where $(\theta_0,n_0)$ is the center of the wave packet
($\langle \hat{\theta} \rangle = \theta_0$, 
$\langle \hat{n} \rangle=n_0$), 
$A$ a normalization constant, and 
$\sigma^{2}=(\Delta n)^2\equiv \langle ( \hat{n} - 
\langle \hat{n} \rangle )^2 \rangle $ 
the variance in the momentum representation \cite{shi}.
We choose $\sigma^2=N/(2\pi L)$ in order to obtain an equal 
value for the variances in $p$ and in $\theta$, 
namely $\Delta \theta \,\Delta p = \hbar_{\text{eff}}$, 
with $\Delta \theta = \Delta p = \sqrt{\hbar_{\text{eff}}}$.
The wave vector (\ref{gauss}) is the closest quantum analog of a 
classical probability density, localized in a small region of the 
phase space, centered in $(\theta_0,p_0)$ and of width $\sigma$.
We point out that, as shown in Ref.~\cite{saraceno}, it is 
possible to prepare efficiently a coherent state on a quantum 
computer.
\item
A \emph{random wave vector}
$|\psi_0\rangle_{\scriptscriptstyle R}=\sum_{n=1}^N c_n |n\rangle$,
where the coefficients $c_n$ have amplitudes of the order
of $1/\sqrt{N}$ (to assure the normalization of the wave vector)
and random phases. This state has no classical analogue.
\end{itemize}

\begin{figure}
\centerline{\epsfxsize=8.5cm \epsffile{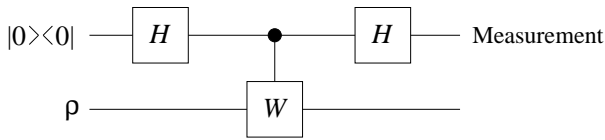}}
\caption{Scattering circuit. The top line denotes a single
ancillary qubit, the bottom line a set of $n_q$ qubits, $H$
the Hadamard gate, and $W$ a unitary transformation.}
\label{fig2}
\end{figure}

The fidelity of quantum motion can be efficiently evaluated
on a quantum computer, as discussed in Ref.~\cite{emerson}.
Here we show an alternative method, based on the 
scattering circuit drawn in Fig.~\ref{fig2}
\cite{zoller,miquelpaz}.
This circuit has various important applications in quantum computation,
including quantum state tomography and quantum spectroscopy
\cite{miquelpaz}. It ends up with a polarization measurement of just 
the ancillary qubit. We measure $\sigma_z$ or $\sigma_y$ and the average
values of these observables are 
\begin{equation}
\langle \sigma_z \rangle = \mathrm{Re} [\mathrm{Tr} (\hat{W} \rho)], \quad
\langle \sigma_y \rangle = \mathrm{Im} [\mathrm{Tr} (\hat{W} \rho)],
\label{scatt}
\end{equation}
where $\langle \sigma_z \rangle$ and $\langle \sigma_y \rangle$ are
the expectation values of the Pauli spin operators $\hat{\sigma}_z$ and
$\hat{\sigma}_y$ for the ancillary qubit, and $\hat{W}$ is a unitary operator
acting on $n_q$ qubits, initially prepared in the state $\rho$
(see Fig.~\ref{fig2}).
These two expectation values can be obtained (up to statistical errors)
if one runs several times the scattering circuit.
If we set $\rho= |\psi_0 \rangle \langle \psi_0 |$ and
$\hat{W} = (\hat{U}^t)^\dagger\, \hat{U}_{\epsilon}^t$, it is
easy to see that 
\begin{equation}
f(t) = \vert \langle \psi_0 \vert
(\hat{U}^t)^\dagger\, \hat{U}_{\epsilon}^t
\vert \psi_0 \rangle \vert^2 =
| \mathrm{Tr} ( \hat{W} \rho) |^{2}=
\langle \sigma_z \rangle^2 + \langle \sigma_y \rangle^2.
\end{equation}
For this reason, provided that 
the quantum algorithm implementing $\hat{U}$ is
efficient, as it is the case for the quantum sawtooth map, 
the fidelity can be efficiently computed by means of the 
circuit shown in Fig.~\ref{fig2}.

\section{Results and Discussion} \label{sec.3}

Hereafter we will call $f_c(t)$ and $f_q(t)$ the fidelity
decays induced by classical or quantum errors, respectively.

Let us first consider the fidelity decay $f_c(t)$, obtained 
under fluctuating perturbations in the parameter $k$ of the 
sawtooth map. 
We will show that, under this type of perturbation, the fidelity 
decay exhibits a marked dependence on the simulated dynamics. 
In particular, qualitatively different behaviors are observed 
depending on the chaotic or non-chaotic motion.

We first consider the \emph{quasi-integrable regime} $-4 \leq K \leq 0$.
In this case the sawtooth map behaves, inside the main integrable 
island with fixed point $(\theta,p)=(\pi,0)$ 
(see Fig.~\ref{fig1}) as a harmonic oscillator, 
with characteristic frequency
$\nu_{\scriptscriptstyle K} = \omega_{\scriptscriptstyle K} / 2\pi
= \sqrt{-K} / 2\pi$.
Therefore, in the semiclassical regime the quantum motion
of coherent wave packets residing inside integrable 
islands closely follows the harmonic evolution of the
corresponding classical trajectories.
In the central island this motion has period
$T= 2 \pi / \sqrt{-K}$, while in the outer islands the period is
multiplied by a factor which depends on the order of the corresponding
resonances (for example, the two upper islands in Fig.~\ref{fig1} 
correspond to a second order resonance, and inside them the period
is doubled).

Since the chosen perturbation affects the parameter $K$,
the fidelity $f_c (t)$ is obtained as the overlap  
of two wave packets which move inside an integrable island
with slightly different frequencies. 
In this case, we know \cite{prosen, gregor} that for a static 
perturbation [$\delta K(t)=\delta K$] the centers of 
the two wave packets separate ballistically (linearly in time) 
and a very fast decay of quantum fidelity is expected 
as far as the distance between the centers of the two 
packets becomes larger than their width $\sigma$.
The type of decay is related to the shape of the initial
wave packet. In particular, for a Gaussian wave packet 
a Gaussian decay is expected.
If $\delta \nu_{\scriptscriptstyle K} \equiv
\nu_{\scriptscriptstyle K+\delta K}-\nu_{\scriptscriptstyle K}$ 
denotes the frequency separation 
between perturbed and unperturbed motion, the Gaussian 
decay takes place after a time
$t_s\propto \sigma/\delta \nu_{\scriptscriptstyle K}$. 

\begin{figure}
\centerline{\epsfxsize=8.5cm \epsffile{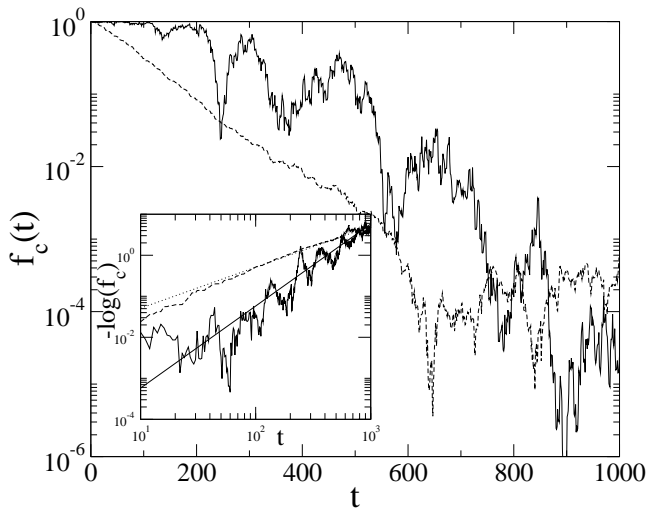}}
\caption{Fidelity decay for the quantum sawtooth map 
with $n_q=12$ qubits, in the presence of a classical
fluctuating perturbation in the $k$ parameter.
The initial condition is a Gaussian wave packet centered
in $(\theta_0,p_0)=(1,0)$.
The upper curve shows the behavior in the quasi-integrable
regime $K=-0.5$, with maximum perturbation strength
$\delta K = 4 \times 10^{-3}$;
the lower one is obtained by simulating the map in the chaotic
regime $K=0.5$, with $\delta K = 2 \times 10^{-4}$.
In the inset we plot the same curves in a graph
showing $-\log(f_c)$ versus time.
The straight lines correspond to exponential fidelity
decay ($-\log f_c\propto t$, upper line) and Gaussian decay
($-\log f_c\propto t^2$, lower line).}
\label{fig3}
\end{figure}

In this paper, we consider the case of a randomly fluctuating 
perturbation $\delta K(t)\in [-\delta K,\delta K]$. 
Therefore, the frequency 
$\nu_{\scriptscriptstyle K+\delta K(t)}$ of a classical
trajectory following the perturbed dynamics is not constant.
The relative displacement of this orbit with respect to the
one described by the unperturbed dynamics
(with a frequency $\nu_{\scriptscriptstyle K}$)
is approximately given by a Brownian motion. 
The separation between the two orbits is proportional
to the frequency difference $\delta\nu_{\scriptscriptstyle K}$.
In this case the fidelity decay is again Gaussian,
but in general it shows large random fluctuations
from the Gaussian profile (see for example the 
upper curve in Fig.~\ref{fig3}), which depend 
on the noise realization. 
Moreover, the distance between the centers of the two wave
packets grows $\propto \sqrt{\delta\nu_{\scriptscriptstyle K} t}$,
and therefore the Gaussian decay starts after a time 
scale $t_s\propto \sigma^2/\delta \nu_{\scriptscriptstyle K}$.

Moreover, the fidelity decay depends not only on the shape of the 
initial state, but also on its position. Indeed, inside any integrable 
island the frequency's perturbation 
$\delta \nu_{\scriptscriptstyle K}=
\nu_{\scriptscriptstyle K+\delta K}-\nu_{\scriptscriptstyle K} 
\approx \frac{\delta K}{4\pi\sqrt{-K}}$
is independent of the position of the wave packet in phase space.
Since larger orbits imply a larger velocity,
and consequently a larger relative ballistic motion of the two
wave packets, the fidelity drops faster when we move far from 
the center of the integrable islands. This is confirmed by
our numerical data (not shown here).

\begin{figure}
\centerline{\epsfxsize=8.5cm \epsffile{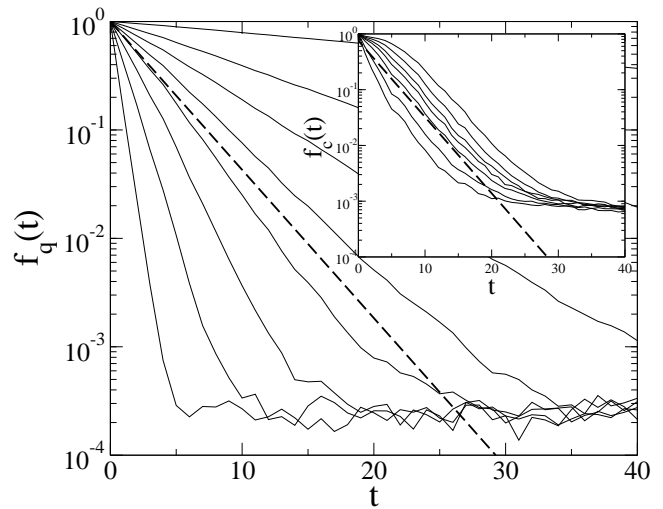}}
\caption{Fidelity decay for noisy gates in the sawtooth map with 
$K=0.1$, $n_q=12$. From right to left: $\epsilon= 1.5 \times 10^{-2},
\, 3 \times 10^{-2}, \, 4 \times 10^{-2}, \, 5 \times 10^{-2},
\, 6 \times 10^{-2}, \, 7.5 \times 10^{-2}, \, 10^{-1},
\, 1.5 \times 10^{-1}$.
Inset: fidelity decay for uncorrelated perturbations in the
parameter $k$. From right to left $\delta K= T\delta k=3 \times 10^{-3},
\, 5 \times 10^{-3}, \, 7.5 \times 10^{-3}, \, 10^{-2}, \, 1.5 \times 10^{-2},
\, 3 \times 10^{-2}, \, 5 \times 10^{-2}$.
In both graphs, data are averaged over 50 initial Gaussian wave packets. 
The two dashed lines show the
Lyapunov exponential decay: $f (t)=e^{- \lambda t}$, where $\lambda \approx
0.315$ is the classical Lyapunov exponent corresponding to $K=0.1$.}
\label{fig4} 
\end{figure}

In the \emph{chaotic regime}, the fidelity $f_c(t)$ always 
decays exponentially, and an example of such decay is 
given in Fig.~\ref{fig3}.
For small perturbations, in the chaotic regime the decay rate 
$\Gamma \propto (\delta K)^2$, as predicted by the Fermi 
golden rule.
However, if the perturbation is strong enough, the fidelity decay
follows a semiclassical regime, in which the decay rate is 
perturbation independent and equal 
to the Lyapunov exponent of the underlying classical 
dynamics (see inset of Fig.~\ref{fig4}).
The condition to observe the Lyapunov decay is that the 
perturbation is quantally strong, namely it couples many
levels ($\delta k> 1$), but classically weak ($\delta k \ll k$).

To summarize, the 
fidelity decay induced by classical perturbations
strongly depends on the dynamical regime, chaotic or
integrable.
The two qualitatively different behaviors (exponential or 
Gaussian decay) are shown in Fig.~\ref{fig3}.
Notice also that the regular dynamics turns out to be
much more stable than the chaotic one (to represent both 
cases on the same figure, the perturbation value chosen 
in the chaotic case is 20 times smaller than the one 
chosen in the integrable case).

\begin{figure}
\centerline{\epsfxsize=7.7cm \epsffile{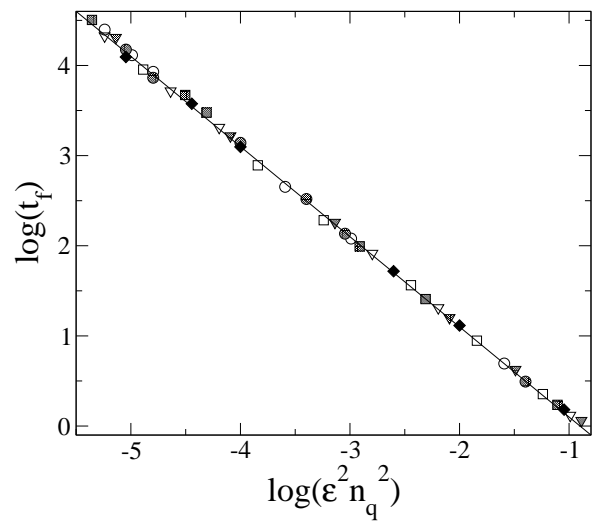}}
\caption{Characteristic time scale $t_f$ for the fidelity decay,
determined by the condition $f(t_f)=0.9$, in the 
sawtooth map at $K=5$, for the case of random noise errors 
in quantum gates.
The data are obtained for different perturbation strengths $\epsilon$
and number of qubits: $n_q=4$ (empty circles), $5$ (filled circles),
$6$ (empty squares), $7$ (filled squares), $8$ (empty triangles),
$9$ (filled triangles), and $10$ (filled diamonds).
The straight line shows the dependence
$t_{f} \simeq 0.126 /\epsilon^{2} n_{q}^{2}$,
corresponding to the exponential fidelity decay (\ref{fidelitydecay}),
with $C\approx 0.28$.
The initial state is in all cases a Gaussian wave packet
and data are averaged over 50 noise realization.}
\label{fig5}
\end{figure}

We now analyze the fidelity behavior in the presence of 
natural errors for quantum computation, namely \emph{random unitary
perturbations} of amplitude $\epsilon$ on \emph{quantum gates}, 
following the noise model described in Sec.~\ref{sec.2}.

As sown in Figs.~\ref{fig4}-\ref{fig5}, 
in the \emph{chaotic regime} the fidelity $f_q (t)$
drops exponentially, with a rate $\Gamma\propto \epsilon^2 n_q^2$
\cite{saturation}. 
This decay can be understood from the Fermi golden rule: 
each noisy gate transfers a probability of order $\epsilon^{2}$ from 
the ideal unperturbed state to other states. Due to the fact that
perturbations acting on two different gates are completely 
uncorrelated, an exponential decay rate proportional to
$\epsilon^{2}$ and to the number of gates $n_g=3n_q^2+n_q$
required to implement one step of the sawtooth map is expected:
\begin{equation}
f_q (t) \simeq e^{-\Gamma t} \simeq e^{-C \epsilon^{2} n_{g} t},
\label{fidelitydecay}
\end{equation}
where $C\approx 0.28$ is a constant which we have computed from 
our numerical data.
We have determined the characteristic time scale $t_f$ for 
fidelity decay from the condition $f_q (t_{f})=A=0.9$
(note that the value chosen for $A$ is not crucial). 
Our numerical calculations, shown in Fig.~\ref{fig5}, 
clearly demonstrate that:
\begin{equation}
t_{f} \propto \frac{1}{\epsilon^{2} n_{q}^{2}},
\label{fidelityscale}
\end{equation} 
in agreement with (\ref{fidelitydecay}).

The fidelity decay in the chaotic regime \emph{always} follows 
the exponential behavior predicted by the Fermi golden rule. 
Therefore, in contrast with 
the case of classical errors, there is no
saturation of the decay rate to the largest Lyapunov 
exponent of the system (see Fig.~\ref{fig4}).

This result can be understood from the \emph{non-locality} of
quantum errors: 
each noisy gate can make direct transfer of 
probability on a large distance in phase space.
This is a consequence of the binary encoding of 
the discretized angle and momentum variables.
For instance, we represent the momentum eigenstates 
$\vert n\rangle$ ($-N/2\le n <N/2$) in the computational 
basis as $\vert \alpha_{n_q} \cdots \, \alpha_2 \alpha_1\rangle$,
where $\alpha_j\in \{0,1\}$ and 
$n=-N/2+N\sum_{j=1}^{n_q} \alpha_j 2^{-j}$.
If we take, say, $n_q=6$ qubits ($N=2^6=64$), the state 
$\vert 000000 \rangle$ corresponds to $\vert n=-32 \rangle$ 
($p=-\pi$), 
$\vert 000001 \rangle$ to $\vert n=-31 \rangle$ 
($p=-\pi+2\pi(1/2^6)$), and so on until  
$\vert 111111 \rangle$, corresponding to $\vert n=31 \rangle$ 
($p=-\pi+2\pi(63/2^6)$). 
Let us consider the simplest quantum error, the bit flip:
if we flip the less significant qubit ($\alpha_1=0\leftrightarrow 1$),
we exchange $\vert n\rangle$ with $\vert n+1\rangle$ 
(mod $N$), while, 
if we flip the most significant qubit ($\alpha_{n_q}=0\leftrightarrow 1$),
we exchange $\vert n\rangle$ with $\vert n+32\rangle$ (mod $N$). 
It is clear that this latter error transfers probability very far 
in phase space and cannot be reproduced by classical local errors.
Therefore, no semiclassical regime for the fidelity decay 
is expected with quantum errors. 
In particular, the non locality of perturbations
makes the fidelity insensitive to the rate of local exponential 
instability, given by the Lyapunov exponent.

\begin{figure}
\centerline{\epsfxsize=8.5cm \epsffile{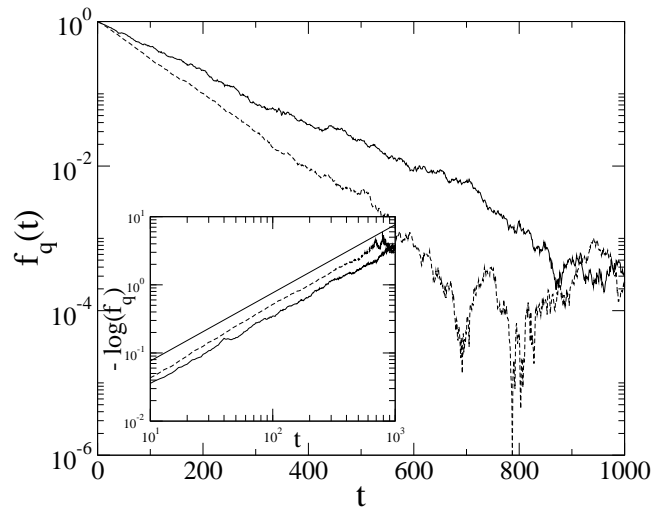}}
\caption{Fidelity decay for the quantum sawtooth map simulated with $n_q=12$
qubits, in the presence of uncorrelated unitary quantum errors with
maximum perturbation strength $\epsilon=10^{-2}$.
As initial condition we consider a Gaussian wave packet
peaked in $(\theta_0,p_0)=(1,0)$.
The upper curve shows the behavior in the quasi-integrable regime $K=-0.5$,
while the lower one is obtained by simulating the map in the chaotic
region $K=0.5$.
In the inset we plot the same curves,  
showing $-\log(f_q)$ versus time.
The solid line corresponds to exponential 
fidelity decay, that is $-\log f_q \propto t$.}
\label{fig6}
\end{figure}

The most striking feature of the fidelity decay induced by quantum 
errors is that it is substantially independent of 
the chaotic or non-chaotic nature of the 
underlying classical dynamics. 
An example of this behavior is shown in Fig.~\ref{fig6}
and strongly contrasts with what obtained by perturbing
the system's parameters (see Fig.~\ref{fig3}).
In particular, the fidelity decay for integrable dynamics 
is exponential, as 
shown in Fig.~\ref{fig6}. If we start from a Gaussian wave packet,
integrable dynamics turns out to be a little more stable than
chaotic dynamics:
we numerically obtained a ratio of the 
decay rates in the chaotic and in the integrable case
which oscillates between $1.15$ and $1.4$, for different 
values of $n_q$ between 5 and 16, and for various 
$\epsilon$ ranging from $10^{-5}$ to $10^{-1}$.

\begin{figure}
\centerline{\epsfxsize=8.6cm \epsffile{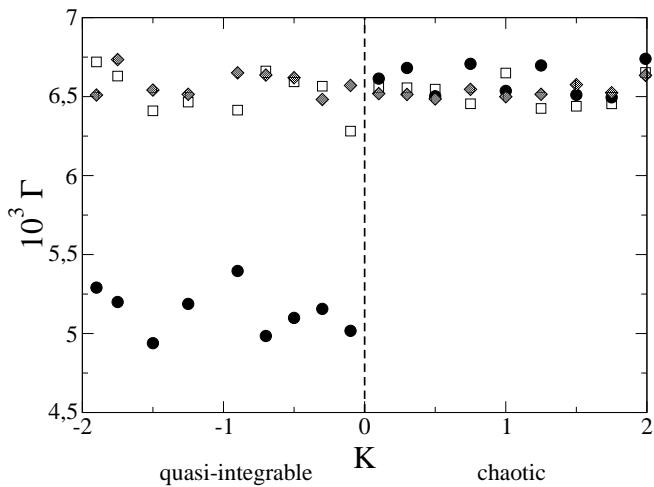}}
\caption{Dependence of the fidelity decay rate,
induced by quantum uncorrelated unitary perturbations,
on $K$, for $n_q=9$, $\epsilon=10^{-2}$.
The dashed line separates the quasi-integrable region
$-4 \leq K \leq 0$ from the chaotic region $K>0$.
As initial condition we choose: (i) a Gaussian wave packet centered
in $(\theta_0,p_0)=(1,0)$ (circles) (note that 
for $-4<K<0$ this packet is inside the main integrable island); 
(ii) a Gaussian packet centered in $(\theta_0,p_0)=(0,0)$ (squares),
that is residing in the diffusive region;
(iii) a random wave function (diamonds).
All data are obtained after averaging
over 25 different noise realizations.}
\label{fig7}
\end{figure}

We stress that the smaller decay rate obtained when we evolve
a Gaussian wave packet inside an integrable island is not
due to the lack of exponential instability but simply to 
the fact that the dynamics preserves the coherence of the 
wave packet.
This can be clearly seen from the data of Fig.~\ref{fig7}.
\begin{itemize}
\item
In the \emph{chaotic regime} $K>0$ (Lyapunov exponent $\lambda>0$), 
the fidelity decay rate is independent of the initial state 
(Gaussian packet or random state) and of the rate of 
exponential instability. Indeed, the decay rate is independent
of $K$, while $\lambda$ depends on  $K$.
\item
In the \emph{quasi-integrable regime} $-4<K<0$ (Lyapunov exponent
$\lambda=0$), only in the case in which we choose as initial 
state a Gaussian packet placed inside an integrable island 
we obtain a fidelity decay rate smaller than in the chaotic 
case. On the other hand, if we start from a random state of if 
we place the Gaussian wave packet inside the anomalously diffusive 
region, we obtain the same decay rate as in the chaotic case. 
\end{itemize}
From these results, we conclude that the decay rate does not 
depend on the value of the Lyapunov exponent.
In short, the decay of the fidelity due to noisy gates is 
\emph{independent of the presence or lack of exponential instability} 
\cite{nodynamics}.
We point out that we have checked that this statement
remains valid also for static errors, like in the case 
in which the dephasing parameters $\epsilon_i,\nu_j$ appearing
in our noise model are time-independent. 

\section{Conclusions} \label{sec.4}

In this paper, we have compared the effects of 
classical and quantum errors on the stability 
of quantum motion.
The main result is that, while the fidelity 
decay under classical errors strongly depends on 
the dynamical nature of the system under investigation 
and on initial conditions, quantum errors act
in a way essentially independent of the system's dynamics.
This practical insensitivity to the dynamics is eventually 
a consequence of the intrinsic non locality of 
the errors that naturally affect the quantum computation.
As a consequence, the rich variety of behaviors found 
in the study of the stability of quantum motion under 
perturbations of the system's Hamiltonian 
\cite{peres,jalabert,pastawski,jacquod,tomsovic,felix,PRE,cohen,prosen,lloyd,baowenli,mirlin,marcos,heller,zurek,geisel,veble,wenge}
has no correspondence in the fidelity of quantum computation 
under its natural perturbations. 
The stability of quantum computation is essentially independent
of the chaotic or integrable behavior of the simulated dynamics.
This conclusion is simply based on the 
non locality of quantum errors and therefore we expect
that it remains valid also in the case of 
non-unitary quantum noise and/or when errors, correlated or 
memoryless, act not only on the qubits on which we apply 
a quantum gate but on all the 
qubits that constitute the quantum computer.

\begin{acknowledgments}

We gratefully acknowledge useful discussions with
Dima Shepelyansky.
This work was supported in part by the EC contracts 
IST-FET EDIQIP and RTN QTRANS, the NSA and ARDA under
ARO contract No. DAAD19-02-1-0086, the PRIN 2002 
``Fault tolerance, control and stability in
quantum information precessing'', and the PA INFM
``Weak chaos: Theory and applications''.

\end{acknowledgments}


\begin{thebibliography}{99}

\bibitem{peres} 
A. Peres, 
Phys. Rev. A {\bf 30}, 1610 (1984).

\bibitem{jalabert} 
R.A. Jalabert and H.M. Pastawski,
Phys. Rev. Lett. {\bf 86}, 2490 (2001).

\bibitem{pastawski} 
F.M. Cucchietti, H.M. Pastawski, and D.A. Wisniacki, 
Phys. Rev. E {\bf 65}, 045206(R) (2002).

\bibitem{jacquod} 
Ph. Jacquod, P.G. Silvestrov, and C.W.J. Beenakker, 
Phys. Rev. E {\bf 64}, 055203(R) (2001),

\bibitem{tomsovic} 
N.R. Cerruti and S. Tomsovic,
Phys. Rev. Lett. {\bf 88}, 054103 (2002).

\bibitem{felix} 
V.V. Flambaum and F.M. Izrailev, Phys. Rev. E
{\bf 64}, 036220 (2001).

\bibitem{PRE} 
G. Benenti and G. Casati, Phys. Rev. E
{\bf 65}, 066205 (2002).

\bibitem{cohen}
D.A. Wisniacki and D. Cohen,
Phys. Rev. E {\bf 66}, 046209 (2002).

\bibitem{prosen} 
T. Prosen, Phys. Rev. E {\bf 65}, 036208 (2002);
T. Prosen and M. \v Znidari\v c,
J. Phys. A {\bf 34}, L681 (2001);
J. Phys. A {\bf 35}, 1455 (2002).

\bibitem{lloyd} 
Y.S. Weinstein, S. Lloyd, and C. Tsallis,
Phys. Rev. Lett. {\bf 89}, 214101 (2002).

\bibitem{baowenli} 
W. Wang and B. Li,
Phys. Rev. E {\bf 66}, 056208 (2002).

\bibitem{mirlin} 
Y. Adamov, I.V. Gornyi, and A.D. Mirlin,
Phys. Rev. E {\bf 67}, 056217 (2003).

\bibitem{marcos} 
I. Garcia-Mata, M. Saraceno, and M.E. Spina,
Phys. Rev. Lett. {\bf 91}, 064101 (2003).

\bibitem{heller} 
J. Van\'\i \v{c}ek and E.J. Heller,
Phys. Rev. E {\bf 68}, 056208 (2003).

\bibitem{zurek} 
F.M. Cucchietti, D.A.R. Dalvit, J.P. Paz, and W.H. Zurek, 
Phys. Rev. Lett. {\bf 91}, 210403 (2003).

\bibitem{geisel} 
M. Hiller, T. Kottos, D. Cohen, and T. Geisel,
Phys. Rev. Lett. {\bf 92}, 010402 (2004).

\bibitem{veble}
G. Veble and T. Prosen,
Phys. Rev. Lett. {\bf 92}, 034101 (2004).

\bibitem{wenge}
W. Wang, G. Casati, and B. Li,
Phys. Rev. E {\bf 69}, 025201 (2004).

\bibitem{schack} 
R. Schack, Phys. Rev. A {\bf 57}, 1634 (1998).

\bibitem{bertrand} 
B. Georgeot and D.L. Shepelyansky,
Phys. Rev. Lett. {\bf 86}, 2890 (2001).

\bibitem{simone}
G. Benenti, G. Casati, S. Montangero, and D.L. Shepelyansky,
Phys. Rev. Lett. {\bf 87}, 227901 (2001).

\bibitem{simone2} 
G. Benenti, G. Casati, S. Montangero, and D.L. Shepelyansky,
Phys. Rev. A {\bf 67}, 052312 (2003).

\bibitem{pomeransky} 
A.A. Pomeransky and D.L. Shepelyansky,
Phys. Rev. A {\bf 69}, 014302 (2004).

\bibitem{libro}
G. Benenti, G. Casati, and G. Strini,
{\it Principles of Quantum Computation and Information},
Vol. 1: Basic Concepts (World Scientific, Singapore, 2004).

\bibitem{cory} 
Y.S. Weinstein, S. Lloyd, J. Emerson, and D.G. Cory,
Phys. Rev. Lett. {\bf 89}, 157902 (2002).

\bibitem{emerson} 
J. Emerson, Y.S. Weinstein, S. Lloyd, and D. Cory,
Phys. Rev. Lett. {\bf 89}, 284102 (2002).

\bibitem{laflamme} 
D. Poulin, R. Blume-Kohout, R. Laflamme, and H. Ollivier, 
Phys. Rev. Lett. {\bf 92}, 177906 (2004).

\bibitem{casati}
G. Casati, B.V. Chirikov, I. Guarneri, and D.L. Shepelyansky,
Phys. Rev. Lett. {\bf 56}, 2437 (1986).

\bibitem{percival} 
I. Dana, N.W. Murray, and I.C. Percival,
Phys. Rev. Lett. {\bf 62}, 233 (1989).

\bibitem{fausto} 
F. Borgonovi, G. Casati, and B. Li, 
Phys. Rev. Lett. {\bf 77}, 4744 (1996).

\bibitem{fausto2} 
F. Borgonovi, 
Phys. Rev. Lett. {\bf 80}, 4653 (1998).

\bibitem{stadium}
G. Casati and T. Prosen, 
Phys. Rev. E {\bf 59}, R2516 (1999).

\bibitem{blatt}
F. Schmidt-Kaler, H. H\"affner, M. Riebe, S. Gulde, 
G.P.T. Lancaster, T. Deuschle, C. Becher, C.F. Roos, 
J. Eschner, and R. Blatt, 
Nature {\bf 422}, 408 (2003).

\bibitem{ciraczoller}
J.I. Cirac and P. Zoller,
Phys. Rev. Lett. {\bf 74}, 4091 (1995)

\bibitem{pazzurek} 
C. Miquel, J.P. Paz, and W.H. Zurek,
Phys. Rev. Lett. {\bf 78}, 3971 (1997).

\bibitem{song}
P.H. Song and D.L. Shepelyansky, 
Phys. Rev. Lett. {\bf 86}, 2162 (2001).

\bibitem{biham}
D. Shapira, S. Mozes, and O. Biham, 
Phys. Rev. A {\bf 67}, 042301 (2003).

\bibitem{bettelli}
S. Bettelli,
Phys. Rev. A {\bf 69}, 042310 (2004).

\bibitem{shi} 
Strictly speaking, the coherent state (\ref{gauss}) should be modified
in order to take into account the periodic boundary conditions 
along $\theta$ and $n$, as explained in S.-J. Chang and K.-J. Shi,
Phys. Rev. A {\bf 34}, 7 (1986). However, this modification does not
significantly affect the results discussed in this paper. 

\bibitem{saraceno} 
J.P. Paz, A.J. Roncaglia, and M. Saraceno, 
Phys. Rev. A. {\bf 69}, 032312 (2004).

\bibitem{zoller} 
S.A. Gardiner, J.I. Cirac, and P. Zoller,
Phys. Rev. Lett. {\bf 79}, 4790 (1997).

\bibitem{miquelpaz} 
C. Miquel, J.P. Paz, M. Saraceno, E. Knill, R. Laflamme,
and C. Negrevergne,
Nature {\bf 418}, 59 (2002).

\bibitem{gregor}
G. Benenti, G. Casati, and G. Veble,
Phys. Rev. E {\bf 68}, 036212 (2003).

\bibitem{saturation} 
The exponential decay stops when the fidelity approaches 
$f_q(\infty)=1/N$. This saturation value is given by the 
inverse of the size of the Hilbert space and reflects
the finiteness of the system.

\bibitem{nodynamics}
We obtained a further confirmation of this statement 
by implementing the quantum
algorithm without dynamical evolution, i.e. by putting 
$T=0$ and $k=0$ in Eq.~(\ref{quantmap}). 
In such a situation, we found that the fidelity drops again 
exponentially and the ratio between the decay rates starting from 
a random or a Gaussian state is the same as in the 
quasi-integrable regime.

\end{thebibliography}

\end{document}